\documentclass[conference]{IEEEtran}
\usepackage{amsmath,amsfonts}
\usepackage{algorithmic}
\usepackage{algorithm}
\usepackage{array}
\usepackage[caption=false,font=normalsize,labelfont=sf,textfont=sf]{subfig}
\usepackage{textcomp}
\usepackage{stfloats}
\usepackage{url}
\usepackage{verbatim}
\usepackage{graphicx}
\usepackage{cite}
\usepackage{booktabs}
\title{ Quantifying Power Systems Resilience Using Statistical Analysis and
Bayesian Learning}

\author{
\IEEEauthorblockN{Apsara Adhikari, Charlotte Wertz, Anamika Dubey}
\IEEEauthorblockA{
Department of Electrical Engineering and Computer Science\\
Washington State University
}
\and
\IEEEauthorblockN{Arslan Ahmad, Ian Dobson}
\IEEEauthorblockA{
Department of Electrical and Computer Engineering\\
Iowa State University
}
}
\begin{document}
\maketitle
\begin{abstract}
The increasing frequency and intensity of extreme weather events is significantly affecting the power grid, causing large-scale outages and impacting power system resilience. Yet limited work has been done on systematically modeling the impacts of weather parameters to quantify resilience.  This study presents a framework using statistical and Bayesian learning approaches to quantitatively model the relationship between weather parameters and power system resilience metrics. By leveraging real-world publicly available outage and weather data, we identify key weather variables of wind speed, temperature, and precipitation influencing a particular region's resilience metrics. A case study of Cook County, Illinois, and Miami-Dade County, Florida, reveals that these weather parameters are critical factors in resiliency analysis and risk assessment. Additionally, we find that these weather variables have combined effects when studied jointly compared to their effects in isolation. This framework provides valuable insights for understanding how weather events affect power distribution system performance, supporting decision-makers in developing more effective strategies for risk mitigation, resource allocation, and adaptation to changing climatic conditions.
\end{abstract}

\section{Introduction}
Climate change is expected to significantly increase both the frequency and severity of extreme weather events \cite{c2es_extreme_weather}. These events pose critical challenges to power grid infrastructure, causing long-duration and widespread outages affecting grid resilience. Weather events such as extreme winds, hurricanes, and heat waves can significantly impact power system resiliency by extending outage duration and increasing economic losses \cite{8966351}. Given the rising frequency and intensity of extreme weather events, the impact on grid resilience and the resulting customer outages is becoming increasingly concerning. The power system plays an essential role in our economy and society as a critical infrastructure. A blackout, for instance, can paralyze transportation systems, disrupt water treatment facilities, and hinder emergency responses \cite{infrastructure_resilience}. Thus, it is imperative to develop systematic approaches to quantify the relationship between weather events and power system resilience, providing actionable insights for planning and preparedness to minimize and mitigate the impacts of future such scenarios. 

Resilience, in the context of power systems, refers to the ability to withstand, adapt to, and recover from disruptions while maintaining or quickly restoring essential operations \cite{resilience}.
There have been several approaches to quantify resilience using existing weather data. In \cite{10373862}, an automated data framework is used to analyze the resiliency of power systems against extreme weather events, along with a spatiotemporal analysis. In \cite{10229392}, machine learning models are proposed to predict outage risk at the state level during and after extreme weather events. In \cite{10012344}, probability density functions are used to describe how resilience metrics are distributed across events. In \cite{he2017nonparametric}, Bayesian learning is used for predictive modeling of storm outages on an electric distribution network. \cite{hossain2019framework} uses a Bayesian network to assess and enhance the resilience of Washington, D.C.’s interdependent electrical infrastructure, identifying reliability, backup power, and resource restoration as key contributing factors.
In \cite{CarringtonPS21},  distribution system resilience is quantified with metrics from utility data.
However, these studies do not quantify the relationship between the attributes of weather parameters (such as severity and intensity) and their impacts on the power grid and grid resilience. In addition, most of the existing literature investigates the effects of a single weather variable on grid resilience. For example, \cite{AhmadPS24} relates the mean area outage rate to wind speed.  
This can obscure the combined effects that different weather parameters can have on the power system. Such models are imperative to conduct predictive analysis to evaluate the impacts and prepare for a future event. Understanding the correlation and complex relationships are crucial, as it provides insights into the dynamics of power system resilience, ultimately forming a foundation for assessing the broader impacts of climate change on grid performance and resiliency.

To address these gaps, this study proposes a framework that integrates historical weather data and outage records to identify patterns and vulnerabilities in power distribution systems. These datasets are publicly available and thus easy for researchers to navigate. By leveraging advanced statistical models and learning techniques, the framework quantifies the relationships between weather parameters and resilience metrics, enabling the development of predictive tools for weather-induced outages. Once these relationships between individual weather variables and resiliency are understood, we then compare this to models that add additional weather variables. 
Essentially, this reveals whether we can gain information when adding more weather parameters than just the most dominant seen in single regression. As a case study, we conduct this weather-resiliency analysis on two counties - Cook County, Illinois, and Miami-Dade County, Florida. These counties are selected because they are both urban populations, but with vastly different weather profiles that could help to understand the regional differences of extreme weather impacts on power grid resilience. The proposed models have diverse applications, including outage forecasting, resilience optimization, and resource allocation, and align with both immediate challenges and long-term climate change adaptation strategies.

\section{Data Description}
\begin{figure*}[h]
    \centering
    \includegraphics[width=0.9\linewidth]{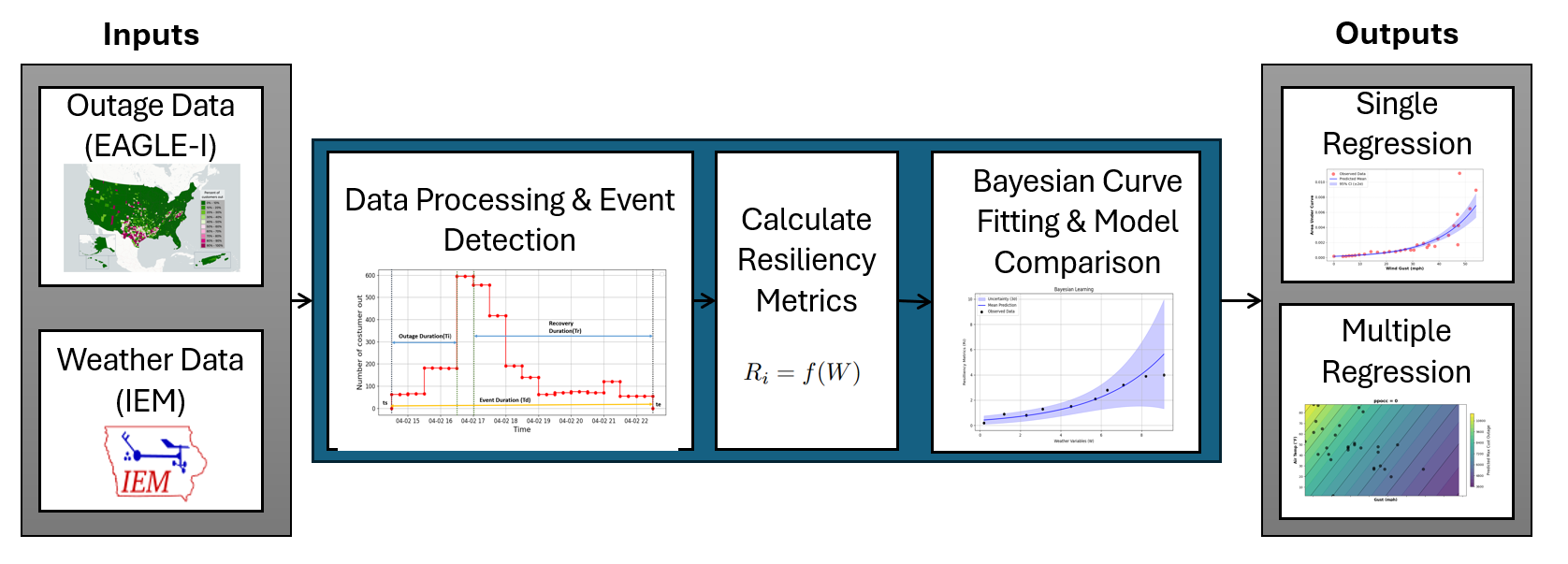}
    \caption{Proposed framework for grid resilience modeling as a function of weather parameters.}
    \label{fig:framework}
\end{figure*}
\subsection{Power Outage Datasets}
The EAGLE‐I dataset \cite{brelsford2024dataset} is an interactive GIS platform developed by Oak Ridge National Laboratory (sponsored by the U.S.\ Department of Energy). It automatically scrapes outage information from utility websites and records county‐level estimates of customers without power at 15‐minute intervals. A few important remarks about the EAGLE-I dataset are worth noting. First, several rural counties exhibit intermittent or missing data due to gaps in utility reporting, motivating our focus on two major urban counties. Second, the 15-minute update interval may result in short-duration outages going unrecorded, introducing imprecision in outage duration estimates. Finally, the dataset reports only the number of customers affected, without identifying individuals, making it difficult to determine whether repeated outage counts correspond to the same or different customers. In this study, we analyze outage data from 2018 through 2023 for two highly urbanized counties: Cook County, Illinois, and Miami–Dade County, Florida. These counties were chosen because the underlying EAGLE-I dataset covers approximately 98\% of electric customers in both Illinois and Florida, ensuring near-complete representation of outage events in these areas.
\subsection{Weather Datasets}
For weather information, we use the Iowa Environmental Mesonet (IEM) Dataset \cite{IEMdatasets}. The IEM maintains an archive of automated airport weather observations from  ASOS (Automated Surface Observing System) and AWOS (Automated Weather Observing System) networks from around the world. 
The IEM dataset comprises a comprehensive suite of meteorological variables such as wind speed, wind gust, precipitation, air temperature, relative humidity, wind direction, and sea‐level pressure. 
In this study, we extract four key meteorological parameters- wind speed, wind gust, precipitation occurrence, and air temperature. 
These weather variables impact electric power outages and we can characterize how system resiliency metrics depend on them. 

\section{Data Processing and Event Extraction}
The weather processing pipeline aggregates meteorological observations from different weather stations across the county, reporting at different intervals to a uniform 15-minute timestamp grid. It retains the maximum value across stations in each bin and imputes missing values via nearest-neighbor interpolation. For the wind‐gust data, any missing observation is interpreted as the absence of a gust and is therefore filled with the concurrent wind‐speed measurement. This is done to enhance the model and collect for each time interval the maximum wind speed reported. The precipitation‐occurrence indicator is modeled as a binary variable. Intervals with missing or zero precipitation depth are assigned a value of zero (no precipitation occurrence), whereas any positive precipitation measurement is coded as one (precipitation occurrence).

The EAGLE-I outage dataset is first filtered by state and then by county to achieve a spatial resolution suitable for regional resilience analysis.
To address missing data and ensure temporal consistency, the dataset is processed using forward fill, where missing values, limited to three consecutive missing entries, are replaced with the last reported observation. 
In addition, the outage records are resampled to uniform 15-minute intervals using a sample-and-hold approach. This ensures that the time series data across all counties follows a regular, consistent timestamp structure, which is critical for aligning with external weather datasets and for resiliency analysis.

\begin{figure}[h!]
    \centering
\includegraphics[width=0.9\linewidth]{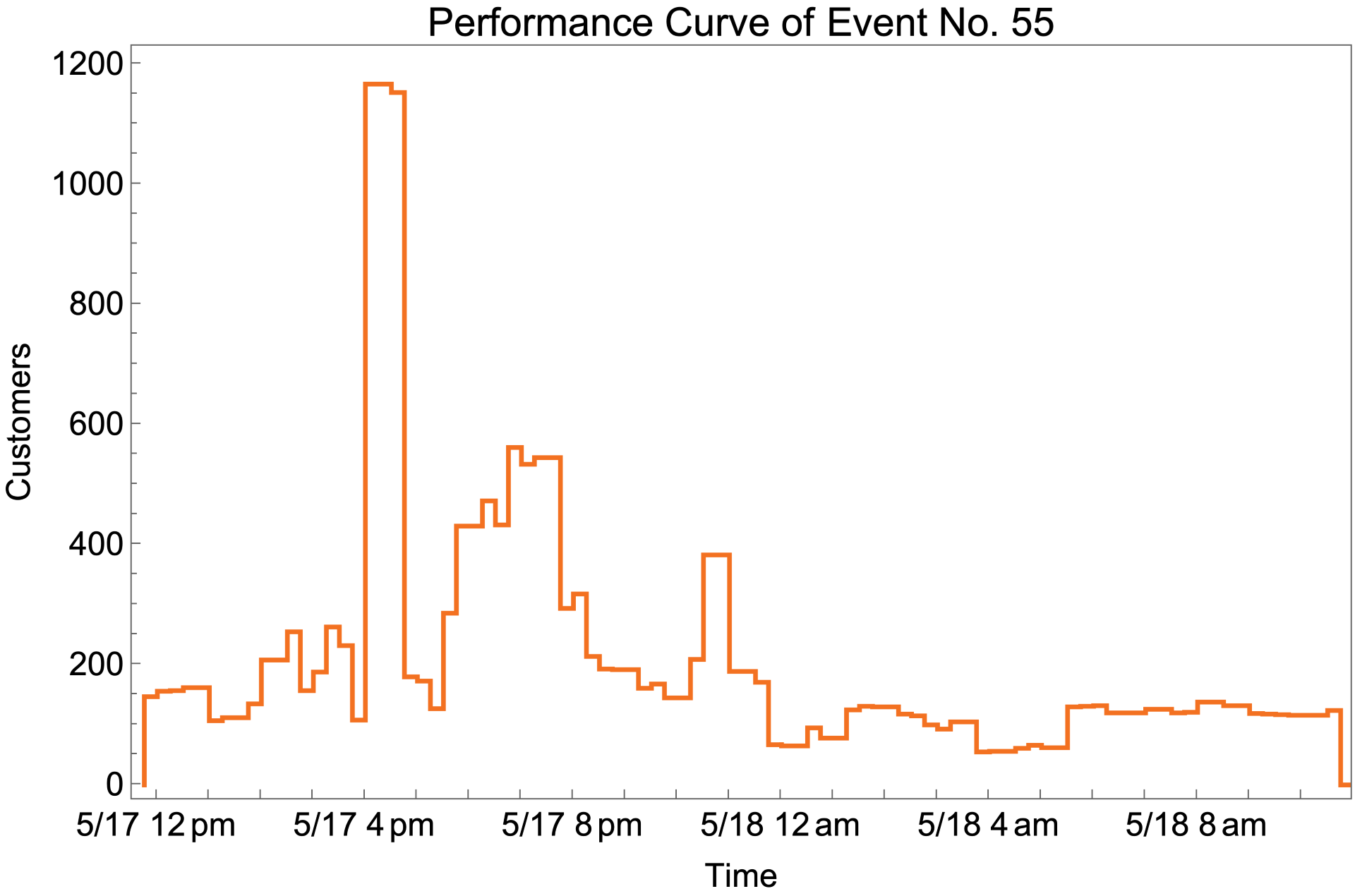}
   \caption{Performance Curve of a sample event}
    \label{PC}
\end{figure}

\looseness=-1
The EAGLE-I records are then aggregated into pre-events using a duration threshold, where consecutive records separated by less than three hours are grouped together as a single pre-event. This ensures that related outage occurrences are grouped together.
The pre-events are further filtered into significant events using a magnitude threshold of 50 customers. This step removes minor outages and filters out minor fluctuations. This two-stage thresholding—based on both duration and magnitude—effectively eliminates low-impact noise and preserves groups of outages likely driven by weather events. 
A significant event is bounded by the  two consecutive times where the number of affected customers passes the magnitude threshold. The start time is  the first time when the outage magnitude exceeds the threshold following a low-outage period, while the end time is the next time when the outage magnitude falls below the threshold, signaling full restoration. 
The performance curve is generated to calculate different resiliency metrics such as area under curve for all distinct events as shown in Figure~\ref{PC}. 
For each significant event, we overlay the full weather time series and compute the peak value of each meteorological variable within that interval to quantify the influence of key weather parameters on the resiliency metrics.

\section{Resiliency Quantification}
Resiliency metrics are used to quantify a power system's ability to withstand and recover from disruptive events.
We use the following resiliency metrics in this work:
\begin{itemize}
    \item \textbf{Area Under the Performance Curve (AUC):} 
    The Area Under the Curve (AUC) serves as a metric for assessing the total customer impact of an outage event. It is defined as the area under the performance curve as well as the total customer hours interrupted in the event \cite{CarringtonPS21}. A lower AUC for a given event indicates higher resilience for a given disruption \cite{10373862}. 

    \item  \textbf{Customer Out:} The total number of customers without electricity during each event  quantifies  the event’s impact. To enable comparisons across different counties, this value is normalized by the total number of electricity consumers in each county, taking into account the coverage of the EAGLE-I dataset.

 
\end{itemize}
By correlating these resiliency metrics with weather conditions, our analysis provides a comprehensive evaluation of the resiliency of the power system. 
For each county, a correlation matrix is generated to evaluate the relationships between weather parameters and resiliency metrics. This matrix helps to identify the dominant weather parameter that has the strongest influence on resiliency metrics. The dominant weather parameters are then selected for learning a single regression Bayesian model based on their significant correlation values.
\begin{figure}[h]
   \centering
   \begin{minipage}{0.48\linewidth}
    \centering
       \includegraphics[width=\linewidth]{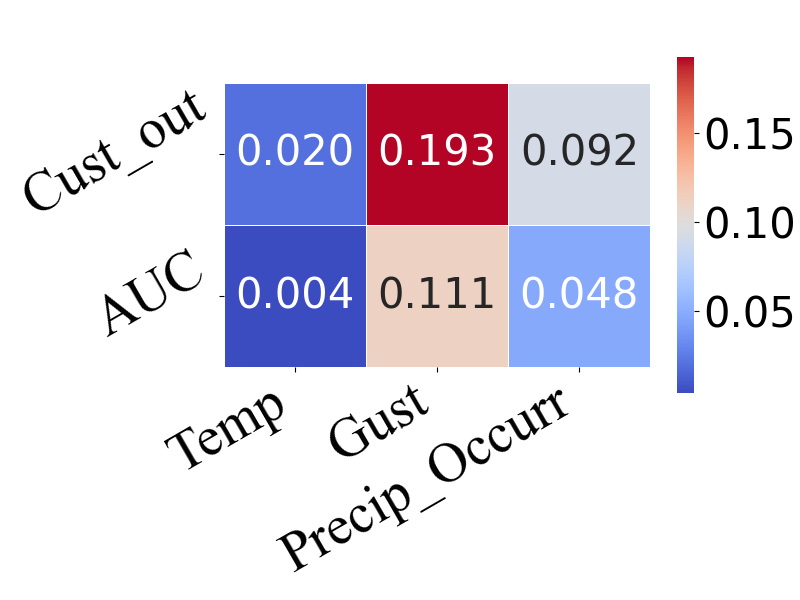}
   \end{minipage}
  \hfill
  \begin{minipage}{0.48\linewidth}
    \centering
   \includegraphics[width=\linewidth]{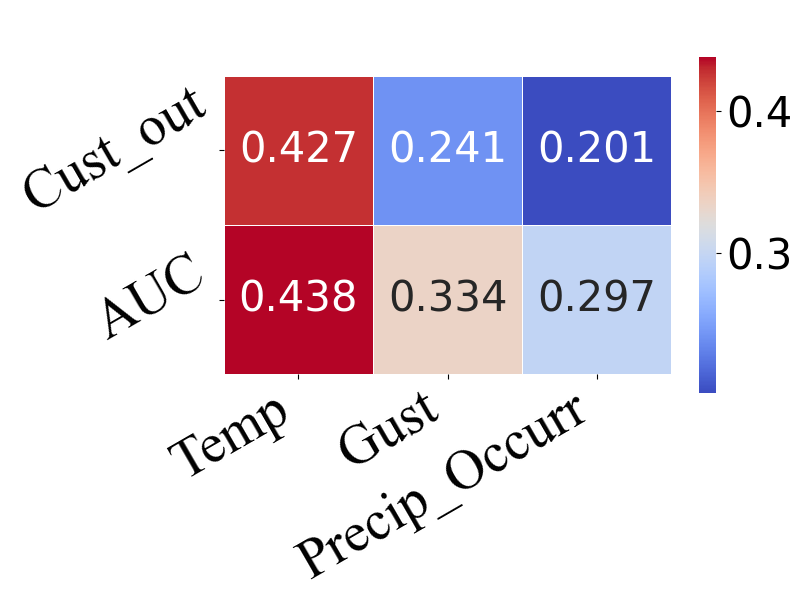}
  
  \end{minipage}
 \caption{Pearson correlation matrix between weather variables and resiliency metrics for Cook County (left) and Miami-Dade County (right).}
 \label{correletation_mat}
\end{figure}
Figure~\ref{correletation_mat} shows the Pearson correlation matrices found from the statistical analysis of the weather parameters compared with resiliency metrics. Here it can be seen that for Cook County, wind gust has a significant impact on resiliency metrics. Conversely, Miami-Dade County has a dual impact from both wind gust and extreme temperatures, with temperature taking the higher correlation. Both relationships are significantly stronger than that of Cook County, likely due to the hurricanes affecting  Florida.

Once the dominant weather parameter has been identified via the correlation analysis, each resiliency metric \( R_i \) is modeled as a function \(R_i = f(W)\) of the individual weather parameters \( W \) using a Bayesian inferential framework. 
%
%
Bayesian learning is employed in this analysis due to its probabilistic framework, which systematically quantifies uncertainty\cite{mazzola1996bayesian}. Given the inherent stochasticity of outage and weather data, Bayesian learning is an effective approach for modeling the complex relationships between weather parameters and resilience metrics. 
In our analysis, the resiliency metrics are assumed to follow an exponential relationship with the weather parameter\footnote{Average outage rate is an exponential function of wind speed in \cite{AhmadPS24}.}, expressed as:
\begin{equation}
R_i = a e^{bW} + c
\end{equation}
where \( R_i \) is the \( i \)-th resiliency metric, \( W \) is the weather parameter (e.g., temperature or wind), \( a, b, c \) are model parameters learned via Bayesian inference, and \( e \) is Euler's constant.

 \looseness-1
After fitting single regression models using the dominant weather predictors, a comprehensive analysis requires evaluating the combined influence of multiple meteorological parameters. Consequently, we employ Bayesian multivariate regression to quantify the joint effects of several weather variables on power grid resiliency. The relationship is formalized as:
\begin{equation}
R_i = f(W_1, W_2, W_3)
\end{equation}
where
\(\displaystyle W_1, W_2, W_3\)  are wind gust, temperature, and precipitation occurrence, and the weather variables significantly contributing to the resiliency metrics in the correlation analysis of Figure \ref{correletation_mat}.
In multivariate regression modeling, it is critical to determine whether the predictors should be incorporated through additive or multiplicative structures. This decision critically influences the model’s capacity to capture complex interactions and nonlinear relationships that may exist among the variables, particularly in the context of resilience modeling influenced by extreme weather conditions.

\begin{figure}[h]
   \centering
   \includegraphics[width=\linewidth]{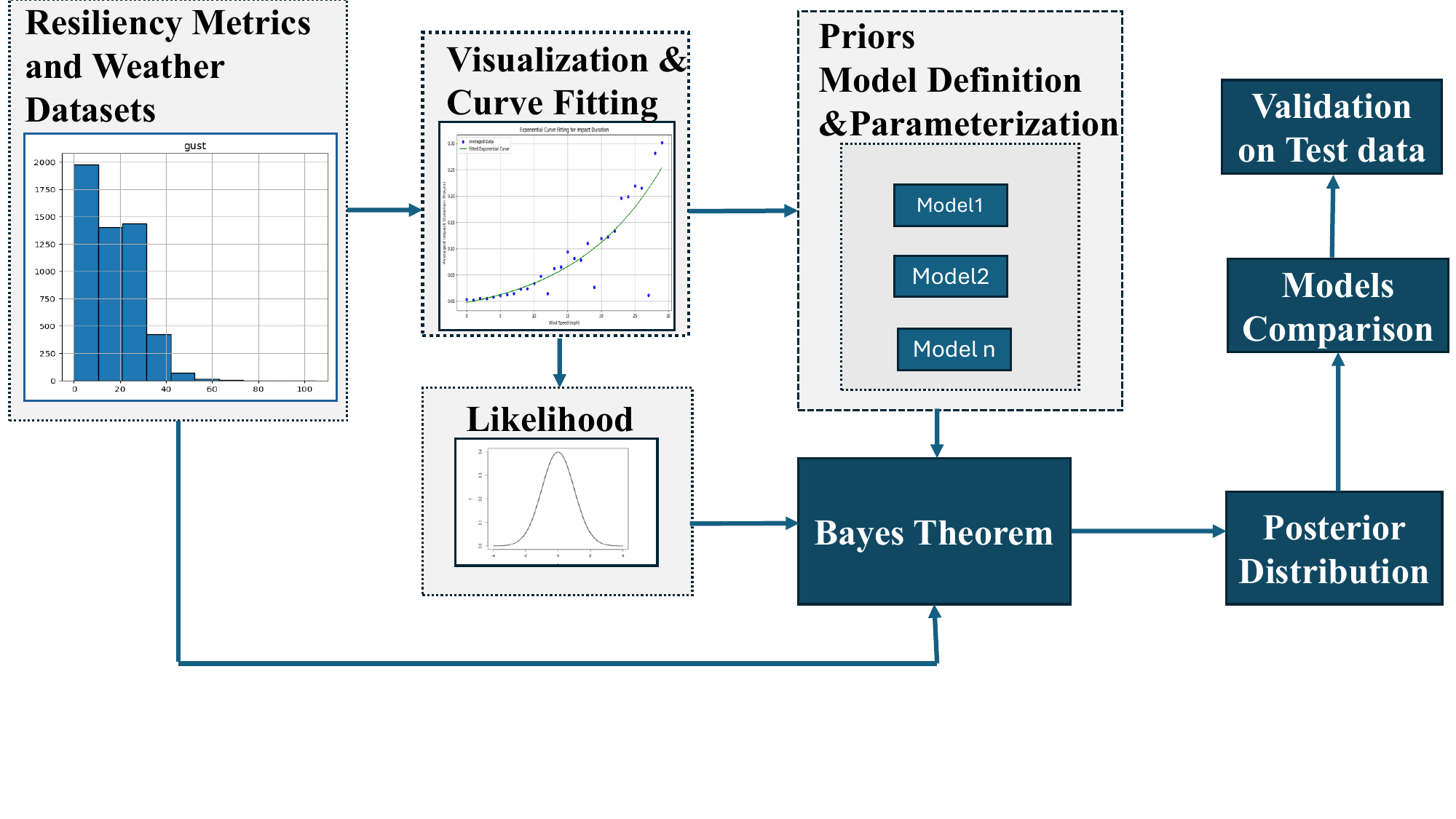} 
   \caption{Bayesian Framework for Multi-Regression Analysis}
   \label{fig:bayesian_framework}
\end{figure}

As illustrated in Figure~\ref{fig:bayesian_framework}, the Bayesian modeling process for multiple variables begins with exploratory data analysis, including visualization and curve fitting that provides valuable insights into the functional relationships between the predictors and the response variable. These insights inform the  model structure and prior distributions, guiding the selection of additive, multiplicative, or interaction-based terms.

Based on this prior understanding, multiple candidate models were formulated and parameterized to capture either additive or multiplicative relationships between predictors. These models were compared against single-variable regression models developed for each respective county. The objective of this comparison is to evaluate the extent to which incorporating additional weather-related predictors enhances model informativeness, while simultaneously considering the trade-off associated with increased model complexity. Each model assumes a normal prior distribution for its parameters and employs Markov Chain Monte Carlo (MCMC) sampling for posterior estimation.

The likelihood function is constructed by quantifying the probability of the observed data under each proposed model structure. Through application of Bayes’ Theorem, the likelihood is integrated with the prior distributions to yield the posterior distribution describing the updated parameter estimates conditioned on the observed data.

Bayes factors are employed to compare competing models by quantifying the relative evidence provided by the data in favor of one model over another, while inherently penalizing excessive model complexity. This approach balances goodness-of-fit with model simplicity, promoting models that are both accurate and efficient. The model exhibiting the highest Bayes factor is selected as the most plausible candidate and subsequently validated using an 80-20 split of the dataset for training and validation. To further assess its generalization capability and predictive accuracy, the selected model is also evaluated on an independent test set.
From the Bayes factor and predictive evaluation metrics shown in Table \ref{bayes_factor} and \ref{tab:model_performance_rms_mae}, we found that the multiplicative model for temperature and wind gust data is more accurate than the additive model. The multiplicative model is as  follows:
\begin{equation}
\label{eq:resilience_model}
R_i \;=\; \exp\bigl(\ln a + b_1 W_1 + b_2 W_2\bigr) \;+\; c
\end{equation}
where \(R_i\) is the \(i\)-th resiliency metric,
\(W_1\) is air temperature,
 \(W_2\) is wind gust,
 \(a, b_1, b_2, c\) are model parameters learned via Bayesian inference, and 
\(\exp\) is the exponential function.
%
The model \eqref{eq:resilience_model} is trained separately for precipitation occurrence and no precipitation, thereby integrating wind gust, air temperature, and precipitation into the prediction of each resiliency metric.
Incorporating multiple meteorological predictors within a single regression framework enables comprehensive characterization of their individual and interactive effects on power system resiliency within a given region.
By including main effects and interaction terms, the model explains how wind gusts, air temperature, and precipitation jointly influence the power system resiliency. Multivariate regression analysis—especially when conducted in a Bayesian framework—facilitates robust decision support for grid adaptation and risk management by quantifying parameter uncertainty and yielding full posterior distributions for predictive quantities. This probabilistic treatment is essential for constructing credible intervals around resilience forecasts and for enabling optimal, risk-aware planning under uncertain weather conditions and system responses.

\section{Results \& Analysis}
\subsection{Single Regression}

For Cook County, the correlation matrix in Figure~\ref{correletation_mat} reveals that wind gust has the highest correlation with resiliency metrics, indicating that it significantly influences the power system's ability to withstand and recover from disruptions. As a result, we model the relationship between wind gust and the resiliency metrics Area Under the Curve (AUC) and normalised Customer Outages. 
For Miami-Dade County, we found that wind gust and temperature are dominant weather parameters, both strongly correlated with the AUC. The resulting model captures the influence of these parameters on the system's performance, which is particularly relevant for Miami-Dade county due to its hot and humid climate, where extreme temperatures and hurricanes can stress the grid \cite{10.1371/journal.pclm.0000523}.

\begin{figure}[h]
    \centering
    \begin{minipage}{0.48\linewidth}
        \centering
        \includegraphics[width=\linewidth]{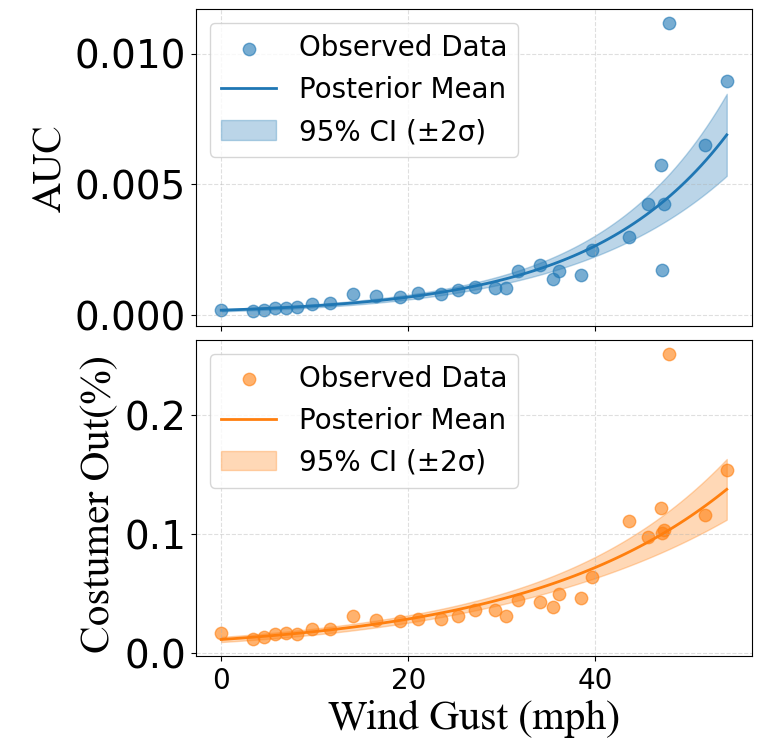}
        \caption{Relationship between Wind gust and Resiliency Metrics for Cook County with 95\% Credible Interval}
        \label{fig:cook}
    \end{minipage}
    \hfill
    \begin{minipage}{0.48\linewidth}
        \centering
        \includegraphics[width=\linewidth]{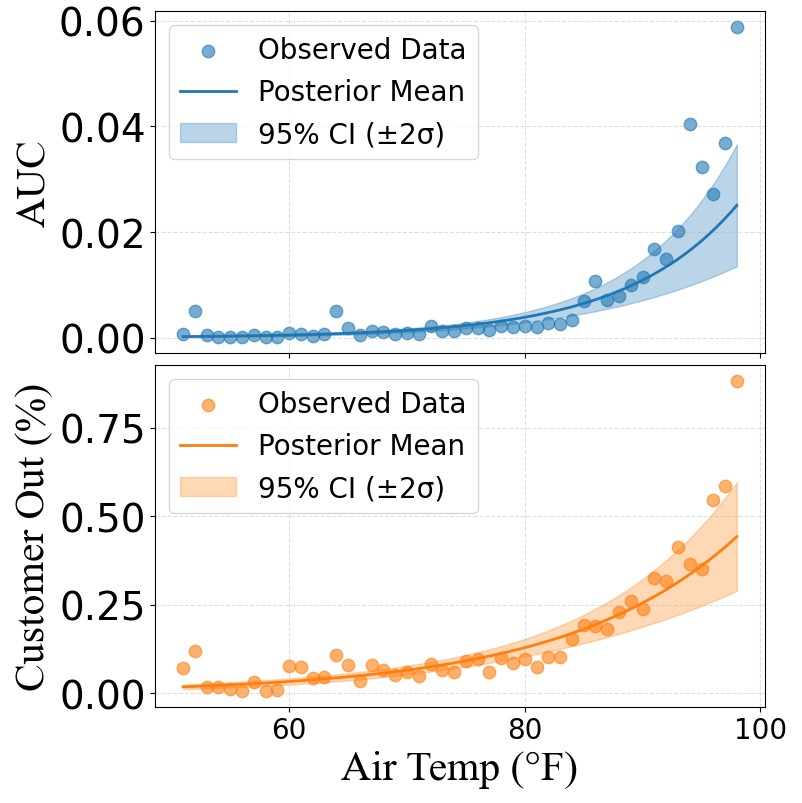}
        \caption{Relationship between Temperature and Resiliency Metrics for Miami-Dade County with 95\% Credible Interval}
        \label{fig:miami}
    \end{minipage}
\end{figure}
\begin{table}[ht]
\centering
\caption{Posterior summaries for Miami–Dade and Cook Counties}
\label{tab:county_params}
\begin{tabular}{lrrrr}
\toprule
\textbf{Parameter} & \multicolumn{2}{c}{\textbf{Miami–Dade}} & \multicolumn{2}{c}{\textbf{Cook}} \\
                   & Mean      & Std       & Mean      & Std       \\
\midrule
$a$                & $0.00411$ & $0.0288$  & $0.0167$  & $0.1177$  \\
$b$                & $0.03943$ & $0.00581$ & $0.06808$ & $0.00363$ \\
$c$                & $-5.1492$ & $6.9940$  & $-4.5666$ & $7.0566$  \\
\bottomrule
\end{tabular}
\end{table}
Table \ref{tab:county_params} reports the Bayesian posterior means and uncertainties for each model parameter in Miami–Dade (temperature‐driven) versus Cook (gust‐driven) fits. In Miami–Dade, the intercept \(a\) is near zero, and the slope \(b \approx 0.039\) (\(\pm 0.0058\)) indicates a moderate positive effect of temperature on the response. In Cook, a larger slope \(b \approx 0.068\) (\(\pm 0.0036\)) confirms that wind gusts have an even stronger and more precisely estimated influence.


\subsection{Multi-Regression Analysis}

Next, additional variables are added to the models using multiple regression. Using the Bayes Factor, we can understand whether this gives us a stronger model than using single regression. We quantify these results in Table \ref{bayes_factor}.
For Cook County, the Bayes factor \(\mathrm{BF}_{\text{gust-only},\,\text{gust+temp}} = 0.178\) implies that the joint gust–temperature model is approximately \(1/0.178 \approx 5.6\) times more probable than the gust-only model, giving “substantial” evidence for combined model. For Miami–Dade County, the Bayes factor \(\mathrm{BF}_{\text{temp-only},\,\text{temp+gust}} = 4.8178\times10^{-13}\) indicates that the combined temperature–gust model is much more likely than the temperature-only model, thus providing “decisive” evidence for the joint predictors model. Incorporating multiple weather variables enhances the model prediction as well, as shown in Table \ref{tab:model_performance_rms_mae}.

\begin{table*}[h!]
\centering
\caption{Model Comparison of Single and Multiple Regression using Bayes Factor}
\label{bayes_factor}
\begin{tabular}{|l|l|l|l|}
\hline
\multicolumn{4}{|c|}{\bfseries Cook County}\\[1.3ex]\hline
\textbf{Model 1}       & \textbf{Model 2}            & {\bfseries\boldmath $\mathrm{\text{Bayes    Factor}}_{1,2}$} & \textbf{Interpretation} \\ \hline
Gust only model       & Gust + Temperature             & 0.178   &    Substantial evidence for Model 2            \\
\hline
\multicolumn{4}{|c|}{\bfseries Miami-Dade County}\\[1.3ex]\hline
\textbf{Model 1}       & \textbf{Model 2}            & {\bfseries\boldmath $\mathrm{\text{Bayes    Factor}}_{1,2}$} & \textbf{Interpretation} \\ \hline

Temperature only model & Temperature + Gust & $4.8178 \times 10^{-13}$ & Strong evidence for Model 2 \\ \hline

\hline
\end{tabular}
\end{table*}


\begin{table}[h!]
\centering
\caption{Model Performance (RMSE and MAE) on Validation and Test Sets for Miame and Cook}
\begin{tabular}{|l|cc|cc|}
\hline
\textbf{Model} & \multicolumn{2}{c|}{\textbf{Validation}} & \multicolumn{2}{c|}{\textbf{Test}} \\
\cline{2-5}
& \textbf{RMSE} & \textbf{MAE} & \textbf{RMSE} & \textbf{MAE} \\
\hline
\multicolumn{5}{|c|}{\textbf{Miami}} \\
\hline
Temperature only     & 0.03701 & 0.037   & 0.0384 & 0.03812 \\
Gust + Temperature   & 0.010   & 0.007   & 0.023  & 0.011 \\
\hline
\multicolumn{5}{|c|}{\textbf{Cook}} \\
\hline
Gust only            & 0.00712 & 0.00655
   & 0.00496
  & 0.00438
 \\
Gust + Temperature   & 0.00656
   & 0.00598
  & 0.00373
  & 0.00302
 \\
\hline
\end{tabular}
\label{tab:model_performance_rms_mae}
\end{table}

For the multiple regression-based approach, both counties are modeled with the wind gust and air temperature variables, as indicated by the results from Table \ref{bayes_factor}. We also filter these by the precipitation binary variable to implicitly add it to the model. We now investigate both models visually. The visualization technique used here is contour plots. For both counties, we plot a contour plot, shaded by the predicted maximum customer outage (normalized by the total number of customers in the county), with wind gust and temperature on the x and y axes, respectively. 
These are plotted for precipitation and no-precipitation cases.

\begin{figure}[h]
    \centering
   \includegraphics[width=\linewidth]{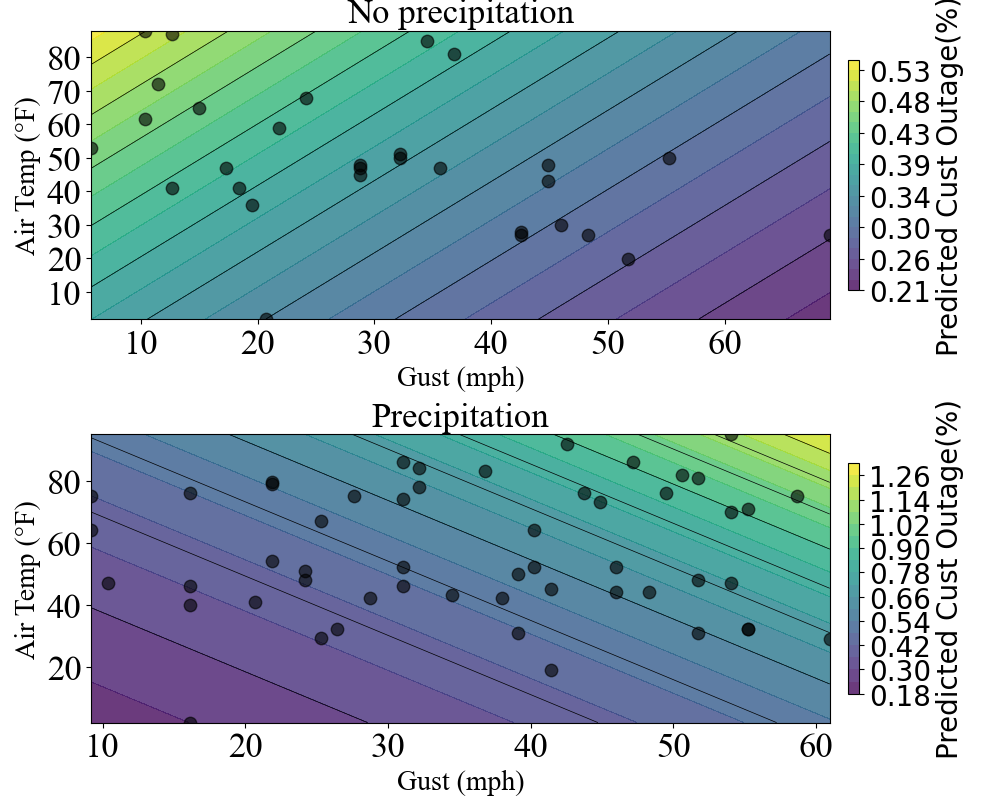}
   \caption{Contour plot for Cook County with Air temperature and wind gust as weather variables filtered by precipitation with customer outage (\%) as resiliency metrics. Black dots indicate the original data points used in model.}
    \label{fig:3variables-Cook}
\end{figure}

Figure \ref{fig:3variables-Cook} shows the wind gust/temperature relationship in Cook County with and without precipitation. Interestingly, when there is no precipitation present, outages are characterized more by extreme temperatures than wind gust. Meanwhile, when there is precipitation, we see far more outages from the dual effects of both wind gust and extreme air temperatures. 
This illustrates the importance of modeling multiple variables rather than single regression because it can adequately capture the type of extreme events causing major power outages. The correlation matrix only tells us which variables have the highest relationship on outage metrics; although in Cook county, wind gust has a dominant and clear relationship on outages, these plots show how air temperature and precipitation further exacerbate the effects of wind gusts.

\begin{figure}[h]
    \centering
   \includegraphics[width=\linewidth]{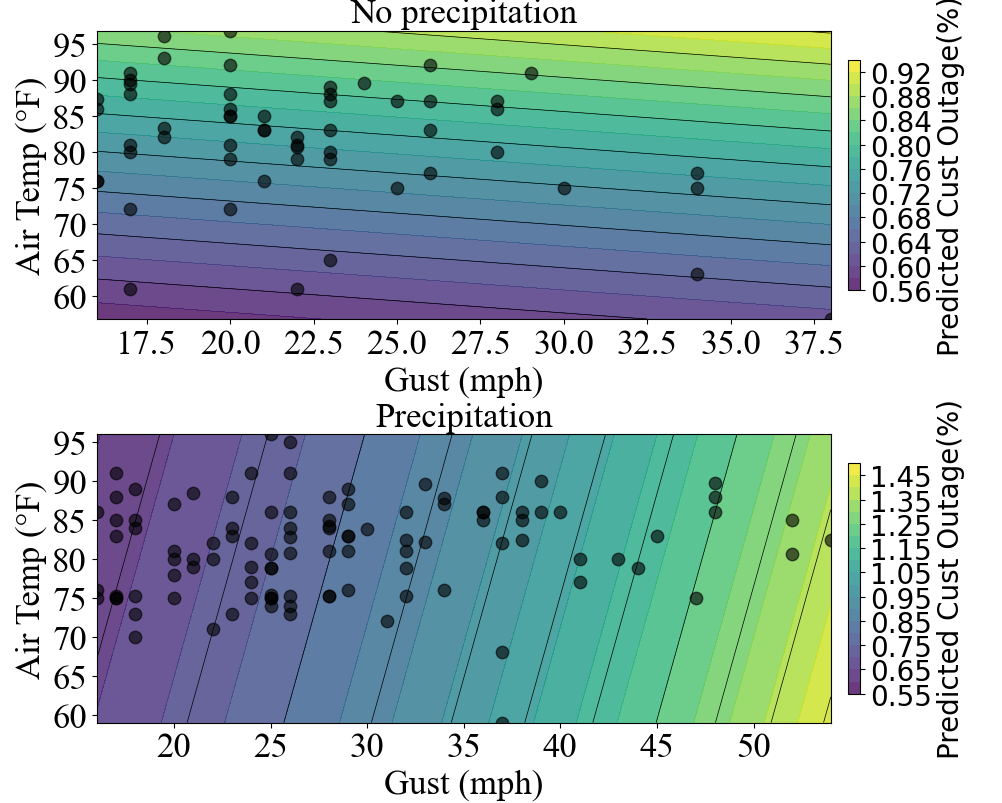}
   \caption{Contour plot for Miami-Dade County with Air temperature and wind gust as weather variables filtered by precipitation with customer outage (\%) as resiliency metrics. Black dots indicate the original data points used in the model.}
    \label{fig:3variables_miame}
\end{figure}

Now, we compare this to the contour mapping for Miami-Dade county in Figure~\ref{fig:3variables_miame}. Here, we see outages characterized by very different types of extreme weather events. Without precipitation, unlike in Cook County, we see outages most commonly within the dual effect of both air temperature and wind gust. Meanwhile, with precipitation, outages are mostly just caused by wind gust, but with a less clear relationship. This is likely due to hurricanes impacting the Miami-Dade region regardless of temperature; in this case, precipitation and wind gust combined model the outages from hurricanes.

\section{Conclusion}
This analysis quantifies how weather parameters have a significant relationship with resiliency metrics, as observed in Miami-Dade County and Cook County. Bayesian analysis was employed to model these relationships, providing a probabilistic framework that captures uncertainty in model parameters. We find from single regression analysis that in Miami-Dade County, wind, temperature, and precipitation have high correlation with outages, while in Cook County, wind alone has a notable impact. However, when weather variables are modeled together, they can capture hidden relationships between weather and power systems resilience that are invisible in single regression analysis, highlighting its importance for risk assessment. These findings underscore the importance of weather parameters in influencing power system resilience. Future work can include expanding the analysis to cover various regions across the United States, including rural regions that are often most impacted by power outages caused by extreme weather. This framework is expected to provide valuable insights for planners and system operators, enabling them to better anticipate and prepare for weather-induced disruptions by informing risk mitigation strategies, resource allocation, and grid adaptation efforts.


\section*{Acknowledgement}
Funding from the Power Systems Engineering Research Center (PSERC) Project S110 is gratefully acknowledged.

\bibliographystyle{IEEEtran}
\bibliography{Bibliography.bib}

@online{c2es_extreme_weather,
  title        = {Extreme Weather and Climate Change},
  author       = {{Center for Climate and Energy Solutions (C2ES)}},
  year         = {2025},
  url          = {https://www.c2es.org/content/extreme-weather-and-climate-change/},
  
}

@ARTICLE{10373862,
  author={Lee, Sangkeun Matthew and Chinthavali, Supriya and Bhusal, Narayan and Stenvig, Nils and Tabassum, Anika and Kuruganti, Teja},
  title     = {Quantifying the Power System Resilience of the {US} Power Grid Through Weather and Power Outage Data Mapping},
  journal   = {IEEE Access},
  volume    = {12},
  pages     = {5237--5255},
  year      = {2024},
  doi       = {10.1109/ACCESS.2023.3347129}
}

@INPROCEEDINGS{10229392,
 author={Lee, Sangkeun and Choi, Jong and Jung, Gs and Tabassum, Anika and Stenvig, Nils and Chinthavali, Supriya},
  title     = {Predicting Power Outage During Extreme Weather Events with {EAGLE-I} and {NWS} Datasets},
  booktitle = {IEEE 24th Int. Conf. on Information Reuse and Integration for Data Science (IRI)},
  year      = {2023},
  pages     = {211--212},
  doi       = {10.1109/IRI58017.2023.00042}}

@ARTICLE{8966351,
 author={Bhusal, Narayan and Abdelmalak, Michael and Kamruzzaman, Md and Benidris, Mohammed},
  title     = {Power System Resilience: Current Practices, Challenges, and Future Directions},
  journal   = {IEEE Access},
  volume    = {8},
  pages     = {18064--18086},
  year      = {2020},
  doi       = {10.1109/ACCESS.2020.2968586}}

@ARTICLE{10012344,
 author={Abdelmalak, Michael and Cox, Jordan and Ericson, Sean and Hotchkiss, Eliza and Benidris, Mohammed},
  title     = {Quantitative Resilience-Based Assessment Framework Using {EAGLE-I} Power Outage Data},
  journal   = {IEEE Access},
  volume    = {11},
  pages     = {7682--7697},
  year      = {2023},
  doi       = {10.1109/ACCESS.2023.3235615}}

@article{brelsford2024dataset,
   author={Brelsford, Christa and Tennille, Sarah and Myers, Aaron and Chinthavali, Supriya and Tansakul, Varisara and Denman, Matthew and Coletti, Mark and Grant, Joshua and Lee, Sangkeun and Allen, Karl and others},
  title     = {A dataset of recorded electricity outages by {U}nited {S}tates county 2014--2022},
  journal   = {Scientific Data},
  volume    = {11},
  number    = {1},
  pages     = {271},
  year      = {2024},
  publisher = {Nature Publishing Group UK London}
}

@article{resilience,
title={On the Definition of Resilience in Systems},
author={Y.Y. Haimes},
journal={Risk Analysis},
volume={29},
pages={498-501},
year={2009},
doi={https://doi.org/10.1111/j.1539-6924.2009.01216.x}}

@article{10.1371/journal.pclm.0000523,
   author = {Do, Vivian AND Wilner, Lauren B. AND Flores, Nina M. AND McBrien, Heather AND Northrop, Alexander J. AND Casey, Joan A.},
  title     = {Spatiotemporal patterns of individual and multiple simultaneous severe weather events co-occurring with power outages in the {U}nited {S}tates, 2018–2020},
  journal   = {PLOS Climate},
  volume    = {4},
  pages     = {1--19},
  year      = {2025},
  month     = {01},
  publisher = {Public Library of Science},
  doi       = {10.1371/journal.pclm.0000523},
  url       = {https://doi.org/10.1371/journal.pclm.0000523}
}

@article{mazzola1996bayesian,
  title={A {B}ayesian approach to managing learning-curve uncertainty},
  author={Mazzola, Joseph B and McCardle, Kevin F},
  journal={Management Science},
  volume={42},
  number={5},
  pages={680--692},
  year={1996},
  publisher={INFORMS}
}

@ARTICLE{AhmadPS24,
  author={Ahmad, Arslan and Dobson, Ian},
  journal={IEEE Trans. Power Systems}, 
  title={Towards Using Utility Data to Quantify How Investments Would Have Increased the Wind Resilience of Distribution Systems}, 
  year={2024},
  volume={39},
  number={4},
  pages={5956-5968},
  doi={10.1109/TPWRS.2023.3342729}}

@ARTICLE{CarringtonPS21,
  author={Carrington, Nichelle’Le K. and Dobson, Ian and Wang, Zhaoyu},
  journal={IEEE Trans. Power Systems}, 
  title={Extracting Resilience Metrics From Distribution Utility Data Using Outage and Restore Process Statistics}, 
  year={2021},
  volume={36},
  number={6},
  pages={5814-5823},
  doi={10.1109/TPWRS.2021.3074898}}

@misc{IEMdatasets,
  author       = {{Iowa State University}},
  title        = {{Iowa Environmental Mesonet: ASOS/AWOS/METAR Data} (IEM Data Portal)},
  howpublished = {\url{https://mesonet.agron.iastate.edu/request/download.phtml?network=FL_ASOS}},

}

@article{infrastructure_resilience,
author = {Natalie Coleman  and Amir Esmalian  and Ali Mostafavi },
title = {Equitable Resilience in Infrastructure Systems: Empirical Assessment of Disparities in Hardship Experiences of Vulnerable Populations during Service Disruptions},
journal = {Natural Hazards Review},
volume = {21},
number = {4},
pages = {04020034},
year = {2020},
doi = {10.1061/(ASCE)NH.1527-6996.0000401},

URL = {https://ascelibrary.org/doi/abs/10.1061/%28ASCE%29NH.1527-6996.0000401},
eprint = {https://ascelibrary.org/doi/pdf/10.1061/%28ASCE%29NH.1527-6996.0000401}}

@article{he2017nonparametric,
  title={Nonparametric tree-based predictive modeling of storm outages on an electric distribution network},
  author={He, Jichao and Wanik, David W and Hartman, Brian M and Anagnostou, Emmanouil N and Astitha, Marina and Frediani, Maria EB},
  journal={Risk Analysis},
  volume={37},
  number={3},
  pages={441--458},
  year={2017},
  publisher={Wiley Online Library}
}

@article{hossain2019framework,
  title={A framework for modeling and assessing system resilience using a Bayesian network: A case study of an interdependent electrical infrastructure system},
  author={Hossain, Niamat Ullah Ibne and Jaradat, Raed and Hosseini, Seyedmohsen and Marufuzzaman, Mohammad and Buchanan, Randy K},
  journal={Int. Journal of Critical Infrastructure Protection},
  volume={25},
  pages={62--83},
  year={2019},
  publisher={Elsevier}
}
\newpage

\end{document}